\documentstyle[11pt,newpasp,twoside]{article}
\begin{document}

\title{Distant Cluster Search: Welcoming some Newcomers from the EIS}
\author{\underline{C. Lobo}, A. Iovino, G. Chincarini}
\affil{Osservatorio Astronomico di Brera - via Brera 28 - 20121 Milano, Italy}

\begin{abstract}
We present a new automated cluster search algorithm, stressing 
its advantages relatively to others available in the literature. 
Applying it to the photometric data of the ESO Imaging Survey 
(EIS, Renzini $\&$ da Costa 1997) allowed us to produce, with a 
high degree of confidence, a new catalogue of cluster candidates 
up to $z \simeq 1.1$.
\end{abstract}


\section{Introduction}
The quest for high--redshift ($z \ga 0.5$) clusters of galaxies has recently 
received a lot of well deserved attention. 
The physical mechanisms that rule galactic evolution still lack a clear 
understanding, and clusters at different redshifts are privileged 
observational targets to develop related studies. Moreover, knowing the number 
density of these systems, as well as their epoch of formation, provides 
crucial ways of testing different theoretical cosmological 
models put forward in the literature (see {\em eg} Bahcall, Fan \& Cen 1997).

\section{The Algorithm}
However, finding clusters at cosmologically interesting look-back times 
($z > 0.5$), not to mention defining a complete sample, is a time consuming 
and difficult task. Successful attempts to gather optically selected samples 
were made by Postman et al (1996; P96 onwards) in the northern hemisphere. 
Prompted by the release of the EIS data, we developed an algorithm - see 
Lazzati et al (1998), Lobo et al (1998) - to 
be applied to catalogues of galaxy positions and magnitudes. One of its main 
advantages is that the spatial and luminosity part of the filter are run 
separately on the catalogue, with no assumption on the typical size or typical 
$M^*$ for clusters, as these parameters intervene in our algorithm only as
typical angular scale - a set of gaussian $\sigma$'s - and typical apparent 
magnitude $m^*$. 
In this way, the significance of a cluster detection is always enhanced, by
combining the most probable $m^*$ with the most probable angular size. 
Moreover, a local background is also used, allowing us to adapt well to and 
overcome the hazards of inhomogeneous data sets (the quality of EIS 
Patch A data was somewhat affected by El Ni\~ no). Extensive simulations using 
directly the EIS data allowed us to obtain completeness and contamination rates 
that do seem advantageous relatively to the respective ones obtained with the 
P96 algorithm (see the next section). 

\section{New Cluster Candidates}
Running our new algorithm on the EIS data produced a new ``robust'' set of 
cluster candidates up to $z \simeq 1.1$ (as estimated {\it via} the respective 
$m^*$), namely $41$ for Patch A ($\sim 3$ sq. degs.) and $21$ for Patch B 
($\sim 1.5$ sq. degs.). Out of these, nearly half are not present in the list 
of candidates obtained from the same data using the P96 algorithm (Olsen et al 
1999; see Lobo et al 1998). The false detection rate we estimated is of 
$\sim 1.3$ spurious candidates per square degree once the threshold is set 
at $S/N = 4$ (our final adopted detection threshold). For
comparison, P96 report an estimated contamination rate for their final
catalogue of at most $30 \%$ in their $5.1$ square degree area. As for 
completeness, setting the detection threshold at $S/N = 4$ still allows 
us to achieve a completeness of $\sim 95\%$ 
until $z \sim 0.9$ for richness 2, Coma-like clusters. 
Some of our cluster candidates have already been 
followed-up in multi-waveband BVRI imaging to estimate photometric redshifts 
and to get a handle on cluster members and determine their color properties.\\
The complete list of candidates, finding charts and further details will 
be presented in Lobo et al (1999, in preparation).

\section{Future}
Three of our highest estimated redshift ($z \ge 0.6$) cluster candidates were 
selected for spectroscopic observations carried out with the FORS/VLT last 
September. The data we obtained (aimed at reaching down to $L \sim 0.3 L^*$, 
and as far as the cluster virial radius) allowed us to confirm their ``cluster 
identity'' and to secure their redshifts as being $z \sim 0.64 $, $z \sim 0.66 
$ and $z \sim 0.71 $, respectively. Once these data are completed, they will 
form a small but homogeneous sample of clusters of very
similar richness and redshift, that can be used to provide us with the absolute
normalization of the bright end of the luminosity function in high--$z$
clusters, an important constraint for theories of structure formation
and evolution.  We will also explore the dynamics and morphological
evolution of member galaxies at a redshift associated to
the assembling of large galaxy clusters.

\acknowledgments
C. Lobo acknowledges financial support by the CNAA fellowship 
reference D.D. n.37 08/10/1997

\end{document}